\documentclass[12pt]{article}
\newcommand{\beq}{\begin{equation}}
\newcommand{\eeq}{\end{equation}}
\newcommand{\beqn}{\begin{eqnarray}}
\newcommand{\eeqn}{\end{eqnarray}}
\newcommand{\bea}[1]{\beq\begin{array}{#1}}
\newcommand{\eea}{\end{array}\eeq}

\newcommand{\Z}{{\cal Z}}
\newcommand{\D}{\displaystyle}
\newcommand{\dual}[1]{{#1}^d}
\newcommand{\diff}{\partial}

\newcommand{\itep}{~\vspace{-1.5cm}\begin{flushright}
{\large ITEP-TH-28/98}\end{flushright}\vspace{1.0cm}}

\begin{document}
\begin{titlepage}
\itep
\begin{center}

{\large\bf Monopole-antimonopole interaction in Abelian Higgs model.}

\vskip 1.5cm

{\bf F.V. Gubarev, M.I. Polikarpov and V.I. Zakharov}

\vskip 1.5cm
{\it Institute of Theoretical and Experimental Physics},\\
{\it 117259 Moscow, Russia}
\vskip 0.5cm
\centerline{and}
\vskip 0.5cm
{\it Max-Planck-Institute for Physics, Werner Heisenberg Institute,}\\
{\it 80805  Munich,
Germany.}\\

\vskip 0.8cm

\end{center}

\begin{abstract}
We consider interaction of a probe monopole-antimonopole pair in the
vacuum of the Abelian Higgs model. For simplicity, the mass of
the Higgs particle is assumed to be much larger
than the mass of the photon (London limit). In case of a massive
photon the straightforward application of the Zwanziger formalism to
accommodate both magnetic and electric charges is known to result in
gauge dependence and infrared instabilities. We argue that the use of
the string representation of the Abelian Higgs model allows to
ameliorate both difficulties. In particular, we arrive at a well
defined expression for the potential energy of the static monopole
sources. We argue that the monopole-antimonopole interaction cannot
be described by a massive photon exchange with a definite propagator
having simple analytical properties.
\end{abstract}
\end{titlepage}

{\bf 1.} Magnetic monopoles emerge with necessity as degrees of
freedom in Abelian projections of gluodynamics \cite{thooft}.
Moreover, there is ample evidence in the lattice simulations for the
relevance and even dominance of the magnetic monopoles in the Maximal
Abelian projection (for recent reviews see, e.g., \cite{polikarpov})
and supporting the idea of the dual superconductor as the confinement
mechanism \cite{nambu}. Hence, there exists the growing interest in
field theoretical approach to description of interaction of monopoles
(see, e.g., papers \cite{kondo} and references therein).

In this note we consider interaction of a static
monopole-antimonopole pair brought to the vacuum of the Abelian Higgs
model. An analogous problem in case of gluodynamics would be
interaction of two static quarks with account of the monopole
condensation. We will not, however, exploit this analogy and
concentrate on the Higgs model itself.

The problem of evaluation of the interaction energy of a monopole
pair in superconductor has its own history ((see
\cite{balachandran,ball,suzuki,suganuma,baker} and references
therein). In the field theoretical approach, the crucial element is
the Zwanziger formalism \cite{zwanziger} which allows to describe
consistently photon interacting with magnetic end electric charges.
In the original formulation of the formalism the vacuum is assumed
to be trivial. If, on the other hand, one tries to incorporate
spontaneous symmetry breaking into the Zwanziger formalism, the resulting
photon propagator appears to be gauge dependent and involves
unphysical singularities. Various ways to deal with these problem based
on physical grounds were proposed \cite{balachandran,suzuki,suganuma}
but the very idea to invoke new principles to derive the propagator
does not look satisfactory from the theoretical point of view. The new
ingredient which we propose is to use the string formulation of the
Abelian Higgs model \cite{akhmedov} which is especially simple if one
considers the London limit, $m_H\gg m_V$.  This approach allows to
evaluate the interaction of the static monopoles in an unambiguous
way. Interpretation of the result in terms of the photon propagator is
not viable, however.

{\bf 2}. Let us start with an overview of the Zwanziger formalism
\cite{zwanziger},  which represents a version of the local field
theory of electrically and magnetically charged particles.  One
introduces two potentials $A_\mu(x)$ and $B_\mu(x)$ which covariantly
interact with electric and magnetic currents, respectively.  The
partition function of the theory reads:

\beq
\Z_{Zw}= \int DA DB \,
e^{-S_{Zw}(A,B)+ie (j^e,A) +ig (j^m, B)}\,   \label{Zw_part_func}
\eeq
where $e$($g$) is the
electric (magnetic) charge, $(j,A) = \int d^4 x\, j_\mu (x) A_\mu
(x) = \int_C A_\mu d \, x_\mu$, and the action $S_{Zw}(A,B)$ is given
by:

\bea{c}\label{Zw-action}
S_{Zw}(A,B)= \int d^4 x\,\{ \frac{1}{2}(n\cdot[\diff\wedge A])^2
+\frac{1}{2}(n\cdot[\diff\wedge B])^2 + \\                                                                          \\
+ \frac{i}{2}(n\cdot[\diff\wedge A])(n\cdot\dual{[\diff\wedge B]}) -
\frac{i}{2}(n\cdot[\diff\wedge B])(n\cdot\dual{[\diff\wedge A]})\},
\eea
where

$$
[A\wedge B]_{\mu\nu} = A_\mu B_\nu - A_\nu B_\mu, \quad
(n \cdot [A\wedge B])_\mu = n_\nu (A\wedge B)_{\nu\mu},
$$
$$
\dual{(G)}_{\mu\nu} =
\frac{1}{2} \varepsilon_{\mu\nu\lambda\rho} G_{\lambda\rho}
$$

The classical equations of motion are:

\bea{r}\label{eq_motion}
(n\diff)^2 A_\mu -n_\mu(n\diff)(\diff A) - \diff_\mu (n\diff) (nA) +
n_\mu \diff^2 (nA)
-i (n\diff) \varepsilon_{\mu\nu\lambda\rho} n_\lambda \diff_\rho A_\nu =
-ie j^e_\mu  \\                                                                                       \\
(n\diff)^2 B_\mu -n_\mu(n\diff)(\diff B) - \diff_\mu (n\diff) (nB)
+ n_\mu \diff^2 (nB)
+i (n\diff) \varepsilon_{\mu\nu\lambda\rho} n_\lambda \diff_\rho B_\nu
= -ig j^m_\mu
\eea
Note that both $j^m$ and $j^e$ are conserved as a consequence of the
equations of motion, $\diff j^{m,e}=0$.

Although the theory contains two gauge fields $A_\mu$ and $B_\mu$, it
still describes one physical photon with two physical degrees of
freedom. This follows from a careful treatment of the Hamiltonian
dynamics of the system~\cite{zwanziger,balachandran}.

We are interested in the interaction energy between classical
external sources $j^m$. Although it may be found directly from the
equations of motion, this is not the easiest way since
(\ref{eq_motion}) is not diagonal in $A$ and $B$ fields. Instead we
integrate out all the fields in~(\ref{Zw_part_func}). First of all,
we have to fix the gauge freedom which is present
in~(\ref{Zw-action}). For the gauge-fixing action we choose

\beq
S_{g.f.}=\int d^4 x\, \left\{\frac{M^2_A}{2} (nA)^2+
\frac{M^2_B}{2} (nB)^2 \right\}
\eeq
and since there is no ghosts for this gauge the
gauge-fixed action of the theory reads:

\bea{c}
S_{Zw}+S_{g.f.}=
\frac{1}{2}\;(A,\hat{D}^{(A)} A) + 
\frac{1}{2}\;(B,\hat{D}^{(B)} B) -
i \;(B,\hat{K} A) -
\\
\\
-ie\;(j^e,A) \; -ig\;(j^m,B)
\eea

In the momentum space $\hat{D}$ and $\hat{K}_{\mu\nu}$ are defined
by:

\bea{c}
\hat{D}^{(A,B)}_{\mu\nu}(k) =\delta_{\mu\nu}(kn)^2-(kn)(n_\mu
k_\nu+n_\nu k_\mu) +(k^2+M^2_{A,B}) n_\mu n_\nu
\\
\\
\hat{K}_{\mu\nu}(k)=(kn)\dual{(n\wedge k)}_{\mu\nu} = (kn)
\varepsilon_{\mu\nu\lambda\rho}n_\lambda k_\rho
\eea

The integration over $B$ field is straightforward yielding the
result:
\bea{c}\label{S(A)}
\Z_{Zw}= \int DA \; e^{\D{- S(A)+ie (j^e, A)}} \\
                                           \\
S(A)=\int d^4 x\, \left\{
\frac{1}{4}(\diff\wedge A - g(n\diff)^{-1}\dual{[n\wedge j^m]})^2 +
\frac{g^2}{2M^2_B} ((n\diff)^{-1} (\diff j^m))^2 + \frac{M^2_A}{2}(nA)^2
\right\}
\eea
Here $(n\diff)^{-1}\dual{(n\wedge j^m)}$ corresponds to the Dirac
string which is parallel to the vector $n$ and is attached to the
current $j^m$.

Performing the last integration over $A$ field we obtain the result:
\bea{c}
\int DA DB \; e^{\D{- S_{Zw}(A,B)+ie (j^e,A) +ig (j^m, B)}} =\;
e^{\D{-S(j^e,j^m)}} \\
\\
S(j^e,j^m)= 
\frac{g^2}{2} \; (j^m,\hat{Q}^{(B)} j^m)+ 
\frac{e^2}{2} \; (j^e,\hat{Q}^{(A)} j^e)+
\frac{i}{4}eg \; (j^e,\hat{T} j^m)
\eea
Thus, the propagators of the original model~(\ref{Zw_part_func})
in the momentum space are:

\bea{c}
<A_{\mu} A_{\nu}>=<B_{\mu} B_{\nu}>=\hat{Q}^{(A,B)}_{\mu\nu}(k)=
\frac{1}{k^2} (\delta_{\mu\nu} +\frac{k^2+M^2_{A,B}}{M^2_{A,B}}\frac{k_\mu k_\nu}{(kn)^2}
-\frac{1}{(kn)}(k_\mu n_\nu+k_\nu n_\mu))
\\
\\
<A_{\mu} B_{\nu}>=\hat{T}_{\mu\nu}=\frac{1}{k^2}\varepsilon_{\mu\nu\lambda\rho}
\frac{n_\lambda k_\rho}{(kn)}=
\frac{1}{k^2 (kn)}([n\wedge k]^d)_{\mu\nu}.
\eea
Note that the last term in the $S(j^e,j^m)$ apparently depends on the
arbitrary chosen vector $n$. The condition that this dependence is
only superficial yields the Dirac quantization condition $eg=4\pi
m$~\cite{zwanziger}. There are general proofs that upon imposing
this condition the physical results are in fact independent on the
choice of the vector $n_{\mu}$. There is a condition attached to the
proofs, namely, that the particle trajectories do not intersect with
the Dirac strings. It might worth noting that the direct use of the
propagators derived above in the basis of plane waves may not satisfy
the last condition and perturbative expressions may depend on
$n_{\mu}$.

{\bf 3}. When the formalism is combined with the spontaneous breaking
of the $U(1)$ symmetry
apparent inconsistencies arise. Indeed, let us assume that a
charged scalar field acquires a nonvanishing vacuum expectation value
or, even simpler, a mass term ${m_V^2\over 2}A_{\mu}^2$ is added to
the Lagrangian.  Then a straightforward diagonalization of the
bilinear terms in the Lagrangian results in the following propagators
of $A$- and $B$-fields (see, e.g., \cite{balachandran}):

\bea{c}\label{bb}
<B_{\mu}
B_{\nu}>(k)~=~ {1\over k^2+m_V^2}\left(\delta_{\mu\nu}+ {m_V^2\over
(kn)^2}(\delta_{\mu\nu}n^2-n_{\mu}n_{\nu})+....\right), \\
\\
<A_{\mu}A_{\nu}>(k)~=~{1\over k^2+m_V^2}\left(\delta_{\mu\nu}+...\right)
\eea
where the dots stand for terms proportional to $k_{\mu}$ and which
can eventually be omitted because of the current conservation.
If we evaluate the interaction energy of a monopole
pair due to the (massive) photon exchange then the
$n_{\mu}$-dependence does not drop off. Also, the double pole in
$(kn)$ causes infrared problems.

All these difficulties arise due to the spontaneous symmetry
breaking. Indeed, in case of the trivial, or perturbative vacuum the 
correction similar to (\ref{bb}) is the first radiative correction to 
$<B_\mu B_\nu>$ propagator due to
a loop of charged scalar particles. In that case the
algebra which led to (\ref{bb}) remains to a great extent unchanged, 
with the following substitutions: the factor $(k^2+m_V^2)^{-1}$ is to 
be replaced by $k^{-2}$ and the factor $m_V^2/(kn)^2$ is to be 
substituted by $k^2/(kn)^2$. As a result, the $n_{\mu}$-dependence of 
the propagator still persists but the overall factor in front of the 
$n_{\mu}$-dependent term is $(kn)^{-2}$. As is readily seen, this
term does not vanish only if the Dirac strings attached to each
monopole are directed along the same line and overlap. The
corresponding energy would be the self energy of the Dirac string and
is unphysical.

The problems with the propagator (\ref{bb}) have been discussed in the
literature \cite{suzuki,suganuma} and prescriptions were proposed to
fix the problems basing on physical arguments. In particular in Ref.
\cite{suzuki} it was proposed to consider the London limit, $m_H\gg
m_V$, take principal value when integrating the pole $(nk)^2$ over
$k$, and to direct $n_{\mu}$ along the line connecting the static
monopole and antimonopole. The reason for these prescriptions is to
imitate the results of the solutions of the classical equations of
motion \cite{ball}. In this note we are looking for a resolution of
the difficulties within the field theoretical approach,
without invoking further hypotheses. It is our understanding that the
difficulties with evaluating, say, static energy of the
monopole-antimonopole pair within the standard formalism are of
principal nature since one cannot avoid overlap of the Dirac strings
and trajectories of charged particles in case of the vacuum
condensation of charges.

{\bf 4}. We propose to utilize the string formulation
\cite{akhmedov} of the Abelian Higgs model (AHM). In more detail, the
partition function of AHM,

\bea{c}\label{AHM}
\Z_{AHM} = \int DA D\Phi D\bar{\Phi} \;
e^{\D{- S_{AHM}(A,\Phi,\bar{\Phi})}}
\\
\\
S_{AHM}(A,\Phi,\bar{\Phi})= \int d^4 x\, \left\{ \;
\frac{1}{4}(\diff\wedge A)^2+ \frac{1}{2} |(\diff-ieA)\Phi|^2 +\lambda
((|\Phi|^2-\eta^2))^2 
\right\}
\eea
in the London limit ($\lambda\to\infty$)
can be exactly rewritten in terms of the word-sheet coordinates
$\tilde{X}(\sigma)$ of the closed ANO strings \cite{akhmedov}:

\beq
\lim\limits_{\lambda\to\infty} \Z_{AHM} =
\int\limits_{\delta\Sigma=0} D\Sigma \; e^{\D{-S(\Sigma)}}
\eeq

The action for the ANO strings reads:

\bea{c}\label{string_representation}
S(\Sigma)=\frac{\pi^2}{e^2}\; m^2_V \;(\Sigma,\hat{K}\Sigma)=
\\
\\
=\frac{\pi^2}{e^2}m^2_V
\int d^2\sigma d^2\sigma' \;
\varepsilon^{ab}\diff_a\tilde{X}^\mu(\sigma) \diff_b\tilde{X}^\nu(\sigma) \;
K(\tilde{X}(\sigma)-\tilde{X}(\sigma')) \;
\varepsilon^{a'b'}\diff_{a'}\tilde{X}^\mu(\sigma') \diff_{b'}\tilde{X}^\nu(\sigma')
\eea
where the kernel $K(x)$ satisfies the equation $(-\diff^2+m^2_V)K(x)=\delta(x)$
and $m^2_V=e^2 \eta^2$.

Consider now the static monopole-antimonopole pair in the vacuum of
the Abelian Higgs model. In the framework of the Zwanziger formalism
this problem corresponds to the expectation value of the Wilson
loop:

\bea{c}
<H(j^m)>= \frac{1}{\Z}\; \int DA DB D\Phi D\bar{\Phi} \;
e^{
-S_{Zw}(A,B)+
\int d^4x\left\{ \;
\frac{1}{2} |(\diff-ieA)\Phi|^2 +
\lambda |(|\Phi|^2-\eta^2)|^2 \right\} \;
+ig \;(j^m,B)
}
\eea
The integral over $B$ is the same as in (\ref{Zw_part_func}), the 
result\footnote{We skip for simplicity the gauge fixing terms.}
is analogous to eq.(\ref{S(A)}) (see also ref.\cite{Akh_dy}):

\bea{c}\label{tHooftAHM}
<H(j^m)>=\frac{1}{\Z_{AHM}} \;
\int DA D\Phi D\bar{\Phi} \;
e^{-S_{AHM}(A,\Phi,\bar{\Phi})}\; H(j^m)
\\
\\
H(j^m)= e^{ \int d^4x \D{\left\{ 
-\frac{1}{4}(\diff\wedge A+\frac{2\pi}{e}\dual{\Sigma_C})^2+
\frac{1}{4}(\diff\wedge A)^2
\right\}}}
\eea
where $\dual{\Sigma_C} = (n\diff)^{-1}\dual{[n\wedge j^m]}$, $\delta
\Sigma_C= j^m$; and we used Dirac quantization condition $e g = 2\pi$.

Thus the expectation value of the Wilson loop $\exp\{ig (j^m,B)\}$ for the
gauge field $B$ is reduced to the expectation value of the 't Hooft loop
$H(j^m)$ for the gauge field $A$. The surface spanned on the loop is
parallel to the vector $n$. Now we show that in the string representation
$<H(j^m)>$ does not depend on the shape of this surface. Consider
for simplicity the London limit, then integrating over the
(nonsingular) phase of the Higgs field and over the gauge field $A$
we have:

\beq\label{tHooftstr}
\lim\limits_{\lambda\to\infty}<H(j^m)> \; =
e^{\D{-\frac{2\pi^2}{e^2}} \D{ (j^m,\hat{K}j^m) }}
\int\limits_{\delta\Sigma=0} D\Sigma \;
e^{\D{-S(\Sigma+\Sigma_c)}}
\eeq
where 
$(j^m,\hat{K}j^m)=\sum_{\cal C,C'}\int_{\cal C}dx_\mu
\int_{\cal C'}dx'_\mu \; K(x-x')$
and the string action $S(\Sigma)$ is the same as
in~(\ref{string_representation}). By the change of variable $\Sigma
\, \to\,  \Sigma -\Sigma_c$ the integral over closed surfaces
($\delta \Sigma = 0$) is reduced to the integral over the surfaces
bounded by the loop $C$ which corresponds to the current $j^m$:

\beq\label{tHooftstring}
\lim\limits_{\lambda\to\infty}<H(j^m)> \; =
e^{\D{-\frac{2\pi^2}{e^2}} \D{ (j^m,\hat{K}j^m) }}
\int\limits_{\delta\Sigma=j^m} D\Sigma \;
e^{\D{-S(\Sigma)}}
\eeq
Thus the dependence on the shape of the Dirac string disappears and
$<H(j^m)>$ depends only on the loop $C$.

{\bf 5}. Now we estimate the static monopole antimonopole potential in the
London limit. For the  monopole-antimonopole pair located at the
distance $R$ one has:
\beq\label{contour}
j_\mu^m(x)=\delta_{\mu,0}[\delta(\vec{x}
- \vec{R}/2)- \delta(\vec{x} + \vec{R}/2)]
\eeq
and the monopole-antimonopole potential is calculated as
\beq
V(R)=-\frac{1}{T}ln<H(j^m)>
\eeq

As for the first term $\frac{2\pi^2}{e^2}\; (j^m,\hat{K}j^m)$
in~(\ref{tHooftstring}) it is easy to find that for the
contour~(\ref{contour}) it gives:

\beq
\frac{2\pi^2}{e^2} (j^m,\hat{K} j^m)
=\mbox{ (self energy) }-\int dt \;
\frac{\pi}{e^2}\; \frac{ e^{-m_V R}}{R}
\eeq
which results in the Yukawa-type
contribution to the potential $V(R)$:

\beq V(R)=V_1(R)+V_2(R),  \qquad
V_1(R)=-\frac{\pi}{e^2} \; \frac{ e^{-m_V R}}{R}
\eeq
As for the second term $V_2(R)$ it is much more involved. Even at the
classical level one has to find the surface bounded by the contour 
$C$ for which the action $S(\Sigma)$ is minimal. It is difficult to 
find such a surface in general case but for the action~(\ref{string_representation})
we expect that it should be the surface of 
minimal area.
For the loop defined by~(\ref{contour}) the minimal surface is the flat surface
parameterized as follows:

\bea{c}
\tilde{X}^0=t \qquad t\in (-\infty ;
+\infty)  \\
\tilde{X}^i=\frac{1}{2}R^i \sigma \qquad \sigma\in (-1 ; +1)
\eea

The calculation of the action $S(\Sigma^{min.}_C)$ is straightforward. Since
for $\Sigma^{min.}_C$:

\beq 
\varepsilon^{ab}\diff_a\tilde{X}^\mu(\sigma)
\diff_b\tilde{X}^\nu(\sigma)=
(\delta^{\mu,0}\delta^{\nu,i}-\delta^{\nu,0}\delta^{\mu,i})\frac{R^i}{2}
\eeq
one has:

\bea{c}
S(\Sigma^{min.}_C)= \frac{1}{2}\left[\frac{\pi R m_V}{e}\right]^2
\int dt dt' d\sigma d\sigma' \; \frac{\D{d^4k}}{\D{(2\pi)^4}} \; \frac{\D{1}}{\D{k^2+m^2_V}} \;
e^{ik(\tilde{X}(\sigma)-\tilde{X}(\sigma'))}= 
\\
\\
=\frac{1}{2}\left[\frac{4\pi R m_V}{e}\right]^2 \int dt \; \frac{\D{d^3 k}}{\D{(2\pi)^3}} \;
\frac{\D{\sin^2(\vec{k}\vec{R}/2)}}{\D{\vec{k}^2+m^2_V}}  \;
\frac{\D{1}}{\D{(\vec{k}\vec{R})^2}}
\eea

Collecting all the above we have for the static monopole-antimonopole
potential:  
\beq 
V(R)=-\frac{\pi}{ e^2} \frac{ e^{-m_V R}}{R} +
\frac{1}{2}\left[\frac{4\pi R m_V}{e}\right]^2 \;\int \frac{d^3 k}{(2\pi)^3}\;
\frac{\sin^2(\vec{k}\vec{R}/2)}{\vec{k}^2+m^2_V} \;
\frac{1}{(\vec{k}\vec{R})^2} 
\eeq

Choosing the Higgs mass $M_H$ as the UV cut-off we finally obtain:
\bea{c}
V(R)=-\frac{\D{\pi}}{\D{e^2}} \;\frac{\D{e^{-m_V R}}}{\D{R}} +
\frac{\pi m^2_V}{2 e^2} \;\left[ \;
R \ln\frac{\D{M^2_H}}{\D{m^2_V}} -\frac{\D{2}}{\D{m_V}} +
\int\limits_{0}^{\infty} dx \;
\frac{\D{e^{-R\sqrt{x+m^2_V}}}}{\D{[x+m^2_V]^{3/2}}}\;\right]
\\
\\
=\frac{\D{\pi m_V}}{\D{2 e^2}} \;\left\{\;
-2 \;\frac{\D{e^{-m_V R}}}{\D{m_V R}} + m_V R \;\ln\left[M^2_H/m^2_V\right]
-2\left[1-e^{-m_V R}\right]
+2 m_V R\mbox{ Ei}[m_V R]
\;\right\}   
\\
\\
\mbox{ Ei}[x]= -\int^{\infty}_{x} \frac{e^{-t}}{t} dt =
\mbox{\bf C}+ \ln[x] + \sum\limits^{\infty}_{k=1} \frac{(-x)^k}{k k!}
\eea
which completes the evaluation of the potential energy of the static
monopole pair in the approximation considered. Note that the analogous
potential was obtained in Ref. \cite{suzuki}, but the additional
regularization was used in this paper.
Moreover our approach is valid beyond the London limit.

{\bf 6}. Thus, we see that the use of the string formulation does allow to
circumvent the difficulties with the photon propagator spelled in
detail above. Namely, there is no infrared divergence since the
expression for the potential contains the combination 
${\sin}^2({\bf k\cdot R})/({\bf k\cdot R})^2$ which is finite if 
${\bf k\cdot R}\rightarrow 0$.
Moreover, the arbitrary vector $n_{\mu}$ of the
Zwanziger formalism is fixed to be directed along the line connecting
the monopoles. This fixation is determined by the minimality
condition of the area of the surface bounded by the world trajectory
of the monopole current.

At first sight we could interpret our results in terms of the photon
propagator as well. Indeed, we could define the static propagator
$D({\bf k}^2)$as:
\beq
D({\bf k}^2)~\equiv~\int{d^3{\bf R}\over (2\pi)^3} V(R)
exp(i{\bf kR}).
\eeq
We would obtain an expression for $D({\bf k}^2)$ without difficulty. Since
we are working in the Euclidean space, the natural guess would be
that in the relativistic version of the propagator we should
substitute ${\bf k}^2$ by $k^2$ and the whole propagator is
$D(k^2)\delta_{\mu\nu} + \frac{k_\mu k_\nu}{k^2} D_1(k^2) $, the last 
term proportional to $k_\mu k_\nu$ is unimportant due to the 
current conservation.

The point is that such a construction would not be valid. Indeed, it
would imply existence of the potential energy of monopole-monopole
interaction proportional to the distance $R$ at large $R$ and of
opposite sign as compared to the case of the monopole-antimonopole
considered above. Such an interaction is devoid of any physical
meaning however. Hence there is no propagator with the standard 
properties. The reason is that the static interaction $V(R)$ considered above 
accounts for the effect of the ANO string which is a classical 
solution to the equations of motion. The classical solutions, on the 
other hand, do not obey the standard crossing properties inherent to 
the standard Feynman propagators.

{\bf 7}. The authors are grateful to R.~Akhoury, M.N.~Chernodub 
and A.~Zhitnitsky for useful discussions. F.V.G. and M.I.P. feel 
much obliged for the kind hospitality extended to them by the staff of 
Max-Planck-Institute for Physics, Werner Heisenberg Institute. This 
work was partially supported by the grants INTAS-RFBR-95-0681, INTAS-96-370 and 
RFBR-96-02-17230a.

\end{document}